\documentclass[aps, onecolumn, superscriptaddress]{revtex4-2}

\usepackage[utf8]{inputenc} 
\usepackage[T1]{fontenc}    
\usepackage{hyperref}       
\usepackage{url}            
\usepackage{booktabs}       
\usepackage{amsfonts}       
\usepackage{nicefrac}       
\usepackage{microtype}      
\usepackage{lipsum}
\usepackage{amsmath}
\usepackage{amsfonts}
\usepackage{amssymb}
\usepackage{indentfirst}
\usepackage{graphicx}
\usepackage{subfigure}
\usepackage{tikz}
\usepackage{pgfplots}
\usepackage{changepage}
\usepackage{xcolor}

\newcommand{\ket}[1]{\ensuremath{\left|#1\right\rangle}}

\newcommand{\matrixel}[3]{\ensuremath{\left\langle #1 \vphantom{#2#3} \right| #2 \left| #3 \vphantom{#1#2} \right\rangle}}

\begin{document}

\title{Perturbation theory for Maxwell's equations in anisotropic materials with shifting boundaries}

\author{Di Yu}
\affiliation{Institute of Fundamental and Frontier Sciences, University of Electronic Science and Technology of China, Chengdu 610054, China}
\affiliation{Yingcai Honors College, University of Electronic Science and Technology of China, Chengdu 611731, China}
\author{Xiaomin Lv}
\affiliation{Institute of Fundamental and Frontier Sciences, University of Electronic Science and Technology of China, Chengdu 610054, China}
\author{Boyu Fan}
\affiliation{Institute of Fundamental and Frontier Sciences, University of Electronic Science and Technology of China, Chengdu 610054, China}
\author{Ju Gao}
\affiliation{Institute of Fundamental and Frontier Sciences, University of Electronic Science and Technology of China, Chengdu 610054, China}
\affiliation{School of Physics, University of Electronic Science and Technology of China, Chengdu 611731, China}
\author{Jingdao Tang}
\affiliation{Institute of Fundamental and Frontier Sciences, University of Electronic Science and Technology of China, Chengdu 610054, China}
\author{Nan Xu}
\affiliation{Institute of Fundamental and Frontier Sciences, University of Electronic Science and Technology of China, Chengdu 610054, China}
\author{You Wang}
\affiliation{Institute of Fundamental and Frontier Sciences, University of Electronic Science and Technology of China, Chengdu 610054, China}
\affiliation{Southwest Institute of Technical Physics, Chengdu 610041, China}
\author{Haizhi Song}
\affiliation{Institute of Fundamental and Frontier Sciences, University of Electronic Science and Technology of China, Chengdu 610054, China}
\affiliation{Southwest Institute of Technical Physics, Chengdu 610041, China}
\author{Qiang Zhou}
\affiliation{Institute of Fundamental and Frontier Sciences, University of Electronic Science and Technology of China, Chengdu 610054, China}
\affiliation{School of Optoelectronic Science and Engineering, University of Electronic Science and Technology of China, Chengdu 610054, China}
\affiliation{CAS Key Laboratory of Quantum Information, University of Science and Technology of China, Hefei 230026, China}
\author{Guangwei Deng}
 \email{gwdeng@uestc.edu.cn}
\affiliation{Institute of Fundamental and Frontier Sciences, University of Electronic Science and Technology of China, Chengdu 610054, China}
\affiliation{CAS Key Laboratory of Quantum Information, University of Science and Technology of China, Hefei 230026, China}

\date{\today}

\begin{abstract}
Perturbation theory is a kind of estimation method based on theorem of Taylor expansion, and is useful to investigate electromagnetic solutions of small changes. By considering a sharp boundary as a limit of smoothed systems, previous study has solved the problem when applying standard perturbation theory to Maxwell's equations for small shifts in isotropic dielectric interfaces\cite{Johnson2002}. However, when dealing with anisotropic materials, an approximation is conducted and leads to an unsatisfactory error. Here we develop a modified perturbation theory for small shifts in anisotropically dielectric interfaces. By using optimized smoothing function for each component of permittivity, we obtain a method to calculate the intrinsic frequency shifts of anisotropic permittivity field when boundaries shift, without approximation. Our method shows accurate results when calculating eigenfrequency's shifts in strong-anisotropy materials, and can be widely used for small shifts in anisotropically dielectric interfaces.
\end{abstract}

\keywords{Perturbation theory \and Anisotropy \and Shifting boundary}

\maketitle

\section{Introduction}
Perturbation theory is an approximation theory for solving eigenvalues and eigenvectors of a characteristic equation based on theorem of Taylor expansion. It is widely used in calculation in quantum mechanics, electrodynamics and other physical fields. 
The classical electrodynamics perturbation theory has great limitations\cite{Johnson1,Johnson2,Johnson3}, including the inability to deal with material boundary movement\cite{Johnson2002,Johnson4,Johnson5,Johnson6}. In fact, the formula for calculating eigenfrequency shifts given by the classical electrodynamics perturbation theory is only suitable for smooth permittivity constant field perturbations and moving metal boundaries\cite{photonic-crystals,Johnson8}. Here, a smooth perturbation on a field means that the overall field shifts smoothly with respect to the change of arguments parameterizing the perturbation. If we use the calculation formula of the intrinsic frequency shifts given by the electrodynamic perturbation theory as a direct generalization without deliberateness, that formula will yield a result including a ill-defined term, which should have been defined at the boundary but unfortunately discontinuous near the boundary\cite{Johnson2002}. If we simply take the value of the term on either side of the boundary and substitute it into the result aforementioned, we will get a wrong result, especially when there is a big dielectric contrast between the two sides of the boundary\cite{Johnson5,Johnson6,Johnson12}. This problem has been solved by Johnson, et al., when the medium in the two sides are both isotropic\cite{Johnson2002}. They treat the value of the permittivity at the boundary (as described by a step function) as the limit of some continuous function. When dealing with anisotropic materials, a common method is to approximately treat them as isotropic materials in order to apply Johnson's formula, such as the approximation on lithium niobate photonic crystal nanocavities\cite{1903,liang}. However, there will be non-negligible errors when using this approximation method, especially when dealing with more complicated calculations, such as the opto-mechanical coupling coefficients.

In this paper, we report a method for calculating the intrinsic frequency shift of anisotropic materials with shifting boundaries. By comparing our method with the previous one in some specific computing tasks, we find that our method is more accurate. What's more, we find Johnson's formula can be derived from our perturbation theory when restricted to the isotropic case. Our method extends the application of Maxwell’s equations to strongly anisotropic materials.

\section{Perturbation theory for the shift of isotropic material boundaries}
There are two requirements to obtain the eigenvalue for the first order perturbation theory: first, the perturbation is smooth; second, the operation in the characteristic equation is Hermitian. The corresponding procedures and requirements for the generalized eigenvalue equation are exactly the same. With these two conditions the perturbation theory can be used to solve the frequency shift of resonant electromagnetic field when the dielectric field is perturbed --- because the time-harmonic electromagnetic field satisfies the following equation\cite{photonic-crystals}:
\begin{equation}\label{1}
\nabla\times\nabla\times\ket{E}=\left(\frac{\omega}{c}\right)^2\left(\frac{\varepsilon}{\varepsilon_0}\right)\ket{E},
\end{equation}
which is a generalized eigenvalue equation. Consequently, one should assume that the perturbation of the dielectric field is smooth to satisfy the first requirement mentioned above. In order to meet the second requirement, we also need to assume that there is no dielectric loss, which leads to a real permittivity. This procedure can also be performed in the case of anisotropic materials, where the dielectric tensor is a real symmetric tensor under the assumption of no loss. In this case, by using a Hermitian operation, we can obtain similar results with isotropic materials, where the only change is that the $\varepsilon$ here is a tensor.

Assuming the perturbation on a given permittivity field is parameterized with an argument $q$, the derivative of intrinsic frequency with respect to $q$ writes
\begin{equation}\label{3}
\frac{d\omega}{dq}=-\frac{\omega^{(0)}}{2}\frac{\matrixel{E^{(0)}}{\frac{d\varepsilon}{dq}}{E^{(0)}}}{\matrixel{E^{(0)}}{\varepsilon}{E^{(0)}}},
\end{equation}
where superscript $(0)$ indicates corresponding initial value without perturbation.

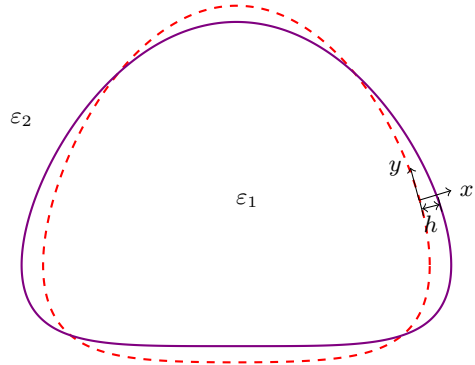
\begin{figure}
\tikzset{elegant/.style={smooth, thick, samples=200}}
\begin{tikzpicture}
\begin{axis}[hide axis,
             xlabel=$x$, ylabel=$f(x)$]
\addplot[elegant, red, dashed, variable=\t, domain=-2*pi:0]
({1.8*(cos(deg(\t)))},{2.2*(sin(deg(\t)))+(sin(deg(\t)))^2});
\addplot[elegant, violet, variable=\t, domain=-2*pi:0]
({2*(cos(deg(\t)))},{2*(sin(deg(\t)))+(sin(deg(\t)))^2});
\node[] at (210,200) {$\varepsilon_1$};
\node[] at (0,300) {$\varepsilon_2$};
\draw[->] (370,200) -- (400,212) node [right] {$x$};
\draw[->] (370,200) -- (362,240) node [left] {$y$};
\draw[|<->|] (372,189) -- node [below] {$h$} (390,195);
\end{axis}
\end{tikzpicture}
\caption{A 2-dimensional schematic diagram showing shifts of boundaries. The permittivities of the materials on each side of the boundary are $\varepsilon_1$ and $\varepsilon_2$ respectively. The local coordinate system is defined to be an orthogonal coordinate system with $x$ axis perpendicular to the boundary. $h$ is the displacement of the boundary in $x$ orientation. These definitions are still available in corresponding 3-dimensional case.}
\label{fig.1}
\end{figure}

We then consider the boundary of two materials that are denoted by their permittivities $\varepsilon_1$ and $\varepsilon_2$ respectively, see Fig.\ref{fig.1}. Obviously, there is an abrupt change of permittivity with the nearby boundary moving, so the perturbation of permittivity is not smooth in this case, which means it is illegal to apply general perturbation here.

To solve this problem, Johnson et al. \cite{Johnson2002} smooth the permittivity field around the boundary to eliminate the abrupt change of the permittivity field accompanying the movement of boundaries so that the change of the permittivity field $\Delta\varepsilon_q (x,y,z)$ is smooth. Then the procedure of perturbation method can be carried out and the approximated shift of eigenfrequency can be obtained. Hence, to get the final result we need to calculate the limitation of Eq. (\ref{3}), namely to calculate the numerator $\matrixel{E^{(0)}}{\frac{d\varepsilon}{dq}}{E^{(0)}}$ and the denominator $\matrixel{E^{(0)}}{\varepsilon}{E^{(0)}}$, respectively. Typically, the denominator $\matrixel{E^{(0)}}{\varepsilon}{E^{(0)}}$ is easy to calculate, while it is much more difficult to calculate the numerator $\matrixel{E^{(0)}}{\frac{d\varepsilon}{dq}}{E^{(0)}}$. Johnson et al. \cite{Johnson2002} calculated this limitation for isotropic materials to be
\begin{equation}\label{4}
\matrixel{E^{(0)}}{\frac{d\varepsilon}{dq}}{E^{(0)}}=\int dA \frac{dh}{dq}[\Delta\varepsilon_{12}|E_{||}^{(0)}|^2-\Delta(\varepsilon_{12}^{-1})|D_{\perp}^{(0)}|^2],
\end{equation}
where the superscript $(0)$ indicates the corresponding initial values without perturbation, the integration domain is the whole boundary. Here $\Delta\varepsilon_{12}=\varepsilon_2-\varepsilon_1$, $\Delta(\varepsilon_{12}^{-1})=\varepsilon_2^{-1}-\varepsilon_1^{-1}$, $|E_{||}^{(0)}|$ means the module of the parallel component for electric field around the border, and $|D_{\perp}^{(0)}|$ denotes the perpendicular component of electric displacement field at the boundary, and $h$ indicates the distance of the movement of infinitesimal boundary patches.

However, this procedure, when conducted on anisotropic materials, will encounter many difficulties because of the complexity of the algebra structure of matrix. To our knowledge, there is no perturbation theory for the case of moving boundaries of anisotropic materials up to now.

\section{The perturbation theory for the shifts of anisotropic material boundaries}
\begin{figure}
\begin{tikzpicture}[
  declare function={
     func1(\x)= (\x < 0)*(1) + (\x > =0)*(2);
     func2(\x)= (\x < -2) * (1)) +  and(\x >= -2, \x < 2) * (0.52*tanh(x)+1.5) + (\x >= 2) * (2);
  }
]
\begin{axis}[
    axis lines = left,
    ymax = 2.5,
    ymin = 0.5,
    xtick = \empty,
    ytick = \empty,
    axis y line = middle,
]
\addplot [
    domain=-5:5,
    samples=500,
    red,
    thick,
]
{func1(x)};
\addlegendentry{$\varepsilon(x)$}
\addplot [
    domain=-5:5,
    samples=100,
    dashed,
    blue,
    thick,
    ]
{func2(x)};
\addlegendentry{$\varepsilon_s(x)$}
\node[] at (960,10) {$x$};
\node[] at (470,185) {$\varepsilon$};
\node[] at (475,7) {O};
\end{axis}
\end{tikzpicture}
\caption{The smoothing from $\varepsilon(x)$ to $\varepsilon_s(x)$, where $s$ parameterizes the smoothing. Obviously, if $x$ changes a little, the change of $\varepsilon_s(x)$ on each position is small, satisfying our smoothing condition.}
\label{fig.2}
\end{figure}
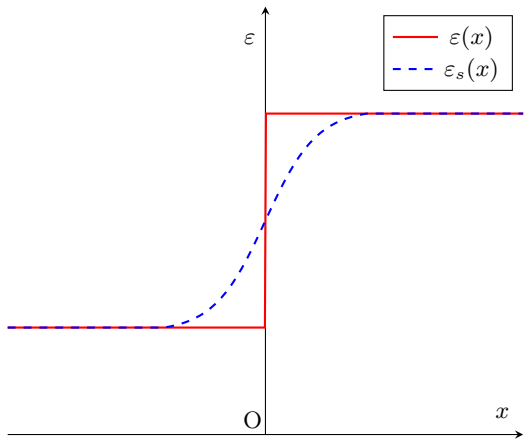

Considering an infinitesimal area of the boundary, we only need to consider its movement in perpendicular orientation to the plane, because the parallel movements of an infinitesimal area does not influence the distribution of the permittivity field. The coordinate system is depicted in Fig.\ref{fig.1}, where we define $x$ as the coordinate in the perpendicular orientation. $x=0$ at the boundary, and $x$ is positive in the $\varepsilon_2$ side. A given perturbation is imposed on the boundary. $h(q)$ is defined to be the change of $x$ coordinate of an infinitesimal area during the perturbation parameterized by argument $q$, and it depends on the location of the infinitesimal area. The unperturbed permittivity distribution is described by $\varepsilon(x)=\varepsilon_1 +(\varepsilon_2-\varepsilon_1)\Theta(x)$, while the perturbed one by $\varepsilon(x)=\varepsilon_1 +(\varepsilon_2-\varepsilon_1)\Theta(x-h)$, where $\Theta(x)=0$ when $x<0$, and $\Theta(x)=1$ when $x>0$. As a result, the perturbation of the permittivity field is $(\varepsilon_2-\varepsilon_1)(\Theta(x-h)-\Theta(x))$, as shown by the red line in Fig.\ref{fig.2}. Because it is not smooth at $x=0$, there will be problems in subsequent calculations. 

To solve this problem, we replace the original permittivity field $\varepsilon(x)$ with a smoothed permittivity distribution function $\varepsilon_s(x)$ and replace $\varepsilon(x-h)$ with $\varepsilon_s(x-h)$, where the argument $s$ parameterizes the smoothing method and $\varepsilon_s(x)\rightarrow \varepsilon(x)$ when $s\rightarrow 0$. In this way the perturbation of the permittivity field becomes smooth. The specific form of $\varepsilon_s(x)$ is undetermined, but we can apply some restrictions on it. Owing to that $\varepsilon_1$ and $\varepsilon_2$ are both real, symmetric, positive definite tensor under lossless assumption, we naturally stipulate the $\varepsilon_s(x)$ to be a real, symmetric, positive definite tensor, which means $\varepsilon_s(x)$ is a Hermitian operation in our calculation. Hence perturbation theory can be applied on the characteristic equation 
\begin{equation}\label{5}
\nabla\times\nabla\times\ket{E}=\left(\frac{\omega}{c}\right)^2 \frac{\varepsilon_s}{\varepsilon_0}\ket{E}.
\end{equation}

The result of the perturbation theory is
\begin{equation}\label{6}
\frac{d\omega}{dq}=-\frac{\omega^{(0)}}{2}\frac{\matrixel{E^{(0)}}{\frac{d\varepsilon_s}{dq}}{E^{(0)}}}{\matrixel{E^{(0)}}{\varepsilon_s}{E^{(0)}}}.
\end{equation}

In order to get the true value of $\frac{d\omega}{dq}$ with no smoothing, we need to calculate the limitation of $\matrixel{E^{(0)}}{\frac{d\varepsilon_s}{dq}}{E^{(0)}}$ and $\matrixel{E^{(0)}}{\varepsilon_s}{E^{(0)}}$ when $s\rightarrow 0$ according to Eq. \ref{6}. While $\matrixel{E^{(0)}}{\varepsilon_s}{E^{(0)}}$ verges $\matrixel{E^{(0)}}{\varepsilon}{E^{(0)}}$ with $s\rightarrow 0$, the limitation of $\matrixel{E^{(0)}}{\frac{d\varepsilon_s}{dq}}{E^{(0)}}$ is elusive. To transform this term to an well-defined form at boundaries for the calculation of the limitation, we now transform it to include only terms continuous at the border. Because $D_x, E_y, E_z$ are all continuous there, it is assumed that the limitation is a function of $D_x, E_y, E_z$, where $x$ axis indicates the orientation of the shifts of the interface.

For convenience, $k_{ij}$ is employed to denote $\frac{d\varepsilon_{s,ij}}{dq}\ (i,j=1,2,3)$, and $E_{ij}$ is employed to denote the matrix with only $(i,j)$ element equal to 1 while others are $0$. In the following discussion, we will omit the field quantity superscript $(0)$. 

To begin with, we have
\begin{equation}\label{7}
[D_x,E_y,E_z]'=E_{11}\varepsilon_s\ket{E}+(E_{22}+E_{33})\ket{E}=A\ket{E},
\end{equation}
where $A=E_{11}\varepsilon_s+E_{22}+E_{33}$, and '$'$' indicates transposition. Here $A$ is reversible, because $\varepsilon_s$ is positive definite tensor with a positive $(1,1)$ element. Hence
\begin{equation}\label{8}
A^{-1}[D_x,E_y,E_z]'=\ket{E}.
\end{equation}
Here $\frac{d\varepsilon}{dh}\neq 0$ only in the domain where the permittivity is smoothed, which is a small area around the boundary. $\Delta h_s$ is defined so that the $x$ coordinate ranges from $-\Delta h_s$ to $\Delta h_s$ in this area. $\Delta h_s$ describe the width of the area with smoothed permittivity, so $\mathop{lim}\limits_{s\rightarrow 0}\Delta h_s=0$. Hence, we have
\begin{equation}\label{9}
\matrixel{E}{\frac{d\varepsilon_s}{dq}}{E}=\int\limits_{R^3}dV[E_x, E_y, E_z]^*\frac{d\varepsilon_s}{dq}[E_x, E_y, E_z]'=\int\limits_{S}dA\frac{dh}{dq}\int_{-\Delta h_s}^{\Delta h_s}[E_x, E_y, E_z]^*\frac{d\varepsilon_s}{dh}[E_x, E_y, E_z]'dx.
\end{equation}
Substituting Eq. (\ref{8}) into Eq. (\ref{9}), we have
\begin{equation}\label{10}
\matrixel{E}{\frac{d\varepsilon_s}{dq}}{E}=\int\limits_{S}dA\frac{dh}{dq}\int_{-\Delta h_s}^{\Delta h_s}[D_x,E_y,E_z]^*A^{-1}\frac{d\varepsilon_s}{dh}A^{-1}[D_x, E_y, E_z]'dx.
\end{equation}
When $s\rightarrow 0$, $\frac{d\varepsilon_s}{dh}=(\varepsilon_1-\varepsilon_2)\delta(x-h),$ where $\delta(x)$ is Dirac function. Because $[D_x, E_y, E_z]$ is continuous at $x=0$, $[D_x, E_y, E_z]^*$  and $[D_x, E_y, E_z]'$ can be taken out of the inner integral, so
\begin{equation}\label{11}
  \matrixel{E}{\frac{d\varepsilon_s}{dq}}{E}=\int\limits_{S}dA\frac{dh}{dq}[D_x,E_y,E_z]^*\int_{-\Delta h_s}^{\Delta h_s}A^{-1}\frac{d\varepsilon_s}{dh}A^{-1}dx[D_x, E_y, E_z]'.
\end{equation}
Now we can only need to ponder how to calculate the limitation of the following term:
\begin{equation}\label{12}
\int_{-\Delta h_s}^{\Delta h_s}Xdx := \int_{-\Delta h_s}^{\Delta h_s}(A^{-1})'\frac{d\varepsilon_s}{dh}A^{-1}dx,
\end{equation}
as other parts in Eq. \ref{11} are already well-defined at the boundary.

Applying Cramer's rule, $A^{-1}$ can be written as
\begin{equation}\label{13}
  A^{-1}=
  \begin{bmatrix}
    \varepsilon_{s,11}^{-1} & -\frac{\varepsilon_{s,12}}{\varepsilon_{s,11}} & -\frac{\varepsilon_{s,13}}{\varepsilon_{s,11}} \\
    0 & 1 & 0 \\
    0 & 0 & 1
  \end{bmatrix}.
\end{equation}
By substituting Eq. (\ref{13}) into Eq. (\ref{12}), we have
\begin{equation}\label{14}
\begin{aligned}
&\int_{-\Delta h_s}^{\Delta h_s}Xdx = \\
&\int_{-\Delta h_s}^{\Delta h_s}
\begin{bmatrix}
  \varepsilon_{s,11}^{-2}k_{11}
  & \varepsilon_{s,11}^{-1}\left(-\frac{\varepsilon_{s,12}}{\varepsilon_{s,11}}k_{11}+k_{12}\right) & \varepsilon_{s,11}^{-1}\left(-\frac{\varepsilon_{s,13}}{\varepsilon_{s,11}}k_{11}+k_{13}\right) \\
  \varepsilon_{s,11}^{-1}\left(-\frac{\varepsilon_{s,12}}{\varepsilon_{s,11}}k_{11}+k_{12}\right) & \varepsilon_{s,12}^2\varepsilon_{s,11}^{-2}k_{11}-2\frac{\varepsilon_{s,12}}{\varepsilon_{s,11}}k_{12}+k_{22} & \varepsilon_{s,12}\varepsilon_{s,13}\varepsilon_{s,11}^{-2}k_{11}-\frac{\varepsilon_{s,13}}{\varepsilon_{s,11}}k_{12}+k_{23} \\
  \varepsilon_{s,11}^{-1}\left(-\frac{\varepsilon_{s,13}}{\varepsilon_{s,11}}k_{11}+k_{13}\right) & \varepsilon_{s,12}\varepsilon_{s,13}\varepsilon_{s,11}^{-2}k_{11}-\frac{\varepsilon_{s,13}}{\varepsilon_{s,11}}k_{12}+k_{23} & \varepsilon_{s,13}^2\varepsilon_{s,11}^{-2}k_{11}-2\frac{\varepsilon_{s,13}}{\varepsilon_{s,11}}k_{13}+k_{33}
\end{bmatrix}
dx.
\end{aligned}
\end{equation}
This is a symmetric matrix, so we only need to consider 6 components of it. We will calculate these components respectively.

In order to calculate the $(1,1)$ element, we consider the smoothing method s. t.
\begin{equation}\label{15}
  \varepsilon_{s,11}^{-1}=\int_{-\infty}^{-\infty} g_s(x-x')\varepsilon_{11}^{-1}dx',
\end{equation}
where $g_s(x)$ is an even smoothing function, and it is non-zero only within a small range with $x$ in $(-\Delta h_{gs}, \Delta h_{gs})$. A smoothing function $S(x)$ is a real function that intensively distributes around $x=0$ and has a unit integral on the whold $x$ axis $\int_{-\infty}^{-\infty}S(x)dx=1$ and verges Dirac function when the related parameter verges 0. According to Eq. (\ref{15}), we have $k_{11}=-\varepsilon_{s,11}^2 \Delta (\varepsilon_{11}^{-1}) g_s(x-h)$, and $\int_{-\Delta h_s}^{\Delta h_s}X_{11}dx=-\Delta(\varepsilon_{11}^{-1})$ for any value of $s$, where $\Delta(\varepsilon_{11}^{-1})=\varepsilon_{1,11}^{-1}-\varepsilon_{2,11}^{-1}$, so $\mathop{lim}\limits_{s\rightarrow 0}\int_{-\Delta h_s}^{\Delta h_s}X_{11}dx=-\Delta(\varepsilon_{11}^{-1})$.

$f_s(x)$ is another smoothing function that is non-zero only within a small range with x in $(-\Delta h_{fs}, \Delta h_{fs})$. An extra limitation imposed the $f_s(x)$ is that $f_s(x)$ verges Dirac function at a lower speed than $g_s(x)$ does when $s\rightarrow 0$, which means $\Delta h_{gs}<< \Delta h_{fs}$ for the same $s$. Taking advantage of $f_s(x)$, we construct the following smoothing method

\begin{figure}
\begin{tikzpicture}[
  declare function={
     func1(\x)= (1/(2*pi*0.5^2)^(1/2))*2.7^(-\x^2/2/0.5^2);
     func2(\x)= (1/(2*pi*1.7^2)^(1/2))*2.7^(-\x^2/2/1.7^2);
     func3(\x)= (\x < -3) * (0)) +  and(\x >= -3, \x <= 3) * (0.167) + (\x > 3) * (0);
  }
]
\begin{axis}[
    axis lines = left,
    xtick = \empty,
    ytick = \empty,
    axis y line = middle,
    ymax = 0.9,
]
\addplot [
    domain=-5:5,
    samples=200,
    red,
    thick,
]
{func1(x)};
\addlegendentry{$g_s(x)$}
\addplot [
    domain=-5:5,
    samples=200,
    dashed,
    blue,
    thick,
    ]
{func2(x)};
\addlegendentry{$f_s(x)$}

\addplot [
    domain=-5:5,
    samples=200,
    green,
    thick,
    ]
{func3(x)};
\addlegendentry{$f_{s0}(x)$}

\node[] at (960,30) {$x$};
\node[] at (470,840) {$y$};
\node[] at (475,30) {$O$};
\end{axis}
\end{tikzpicture}
\caption{A schematic for different smoothing functions. A smoothing function $S(x)$ is a real function that intensively distributes around $x=0$, has a unit integral on the whold $x$ axis $\int_{-\infty}^{-\infty}S(x)dx=1$, and verges Dirac function when the related parameter verges 0. A possible form of $f_s(x)$ and $g_s(x)$ for the same $s$ is shown in this graph, where $g_s(x)$ is much sharper than $f_s(x)$, satisfying the condition we impose. $f_{s0}(x)$ is another possible form of $f_s(x)$ corresponding to rectangular smoothing method, and the parameter s here may represent the width of the rectangle.}
\label{fig.3}
\end{figure}
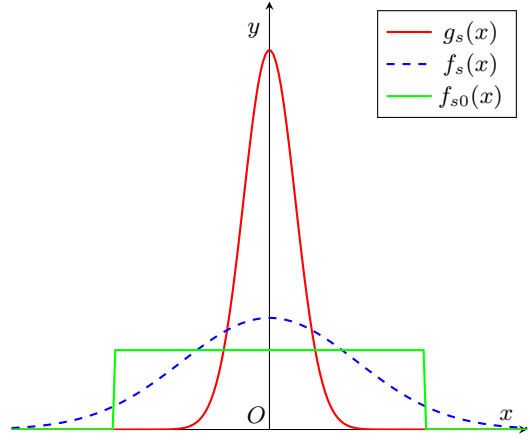

\begin{equation}\label{16}
  \varepsilon_{s,ij}=\int_{-\infty}^{\infty}f_s(x-x')\varepsilon_{ij}dx'.
\end{equation}
For convenience, We define $k_{ij}=\Delta\varepsilon_{ij}f_s(x-h)$, where $i, j\in{1,2,3}$, and $\Delta\varepsilon_{ij}=\varepsilon_{1,ij}-\varepsilon_{2,ij}$. Substituting these equations into $X_{12}, X_{13}$ and considering the parity of $f_s(x)$ and $g_s(x)$, we have
\begin{equation}\label{17}
\begin{aligned}
  &\mathop{lim}\limits_{s\rightarrow 0}\int_{-\Delta h_s}^{\Delta h_s}X_{1k}dx=\mathop{lim}\limits_{s\rightarrow 0}\int_{-\Delta h_s}^{\Delta h_s}X_{k1}dx=\frac{\varepsilon_{1,1k}}{\varepsilon_{1,11}}-\frac{\varepsilon_{2,1k}}{\varepsilon_{2,11}}\ (k=2,3).
\end{aligned}
\end{equation}

For the remaining parts $\int_{-\Delta h_s}^{\Delta h_s}X_{ij},\ i,j=2,3$, each one consists of three terms in terms of Eq. \ref{14}. The last term o which is easy to calculate: $\int_{-\Delta h_s}^{\Delta h_s}k_{ij}dx=\Delta \varepsilon_{ij}$. As for the first term $\int_{-\Delta h_s}^{\Delta h_s}\varepsilon_{s,1i}\varepsilon_{s,1j} \varepsilon_{s,11}^{-2} k_{11}dx$, we can replace $\varepsilon_{s,12}$ and $\varepsilon_{s,13}$ by $\frac{\varepsilon_{1,12}+\varepsilon_{2,12}}{2}$ and $\frac{\varepsilon_{1,13}+\varepsilon_{2,13}}{2}$, respectively, because the non-zero region of $\varepsilon_{s,11}^{-2}k_{11}$ is much smaller than that of $\varepsilon_{s,12}, \varepsilon_{s,13}$. Finally we turn to the second term
\begin{equation}\label{18}
\frac{\varepsilon_{s,1i}}{\varepsilon_{s,11}}k_{1j}\ (i,j=2,3).
\end{equation}

Owing to the high convergence rate of $g_s(x)$, $\varepsilon_{s,11}$ can be seen as a piecewise function taking a uniform value when $x<0$, while taking the other uniform value when $x>0$. Whereupon
\begin{equation}\label{19}
  \int_{-\Delta h_s}^{\Delta h_s}\frac{\varepsilon_{s,12}}{\varepsilon_{s,11}}k_{12}dx = \frac{1}{\varepsilon_{1,11}}\int_{-\Delta h_s}^{0}\varepsilon_{s,12}k_{12}dx + \frac{1}{\varepsilon_{2,11}}\int_{0}^{\Delta h_s} \varepsilon_{s,12}k_{12}dx.
\end{equation}
Noting that $(-\Delta h_{fs}, \Delta h_{fs})=(-\Delta h_{s}, \Delta h_{s})$, we have $\varepsilon_{s,12}|_{x=-\Delta h_s}=\varepsilon_{1,12}$, $\varepsilon_{s,12}|_{x=0}=\frac{\varepsilon_{1,12}+\varepsilon_{2,12}}{2}$, and $\varepsilon_{s,12}|_{x=\Delta h_s}=\varepsilon_{2,12}$. Then it is reasonable for us to approximate $\varepsilon_{s,12}\approx \frac{3}{4}\varepsilon_{1,12}+\frac{1}{4}\varepsilon_{2,12}$, when$ x<h$; $ \frac{1}{4}\varepsilon_{1,12}+\frac{3}{4}\varepsilon_{2,12}$, when $x>h$. Substituting these into Eq. (\ref{19}), we have
\begin{equation}\label{20}
  \mathop{lim}\limits_{s\rightarrow 0}\int_{-\Delta h_s}^{\Delta h_s}\frac{\varepsilon_{s,12}}{\varepsilon_{s,11}}k_{12}dx \approx \frac{1}{8}\left[\frac{3\varepsilon_{1,12}+\varepsilon_{2,12}}{\varepsilon_{1,11}}+\frac{\varepsilon_{1,12}+3\varepsilon_{2,12}}{\varepsilon_{2,11}}\right](\varepsilon_{1,12}-\varepsilon_{2,12}).
\end{equation}
Many specific choice of the smoothing function can satisfy the equation above accurately, so this step would not introduce errors. For example, we can choose rectangular impulse function $f_{s0}(x)$, then we have
\begin{equation}\label{21}
  \mathop{lim}\limits_{s\rightarrow 0}\int_{-\Delta h_s}^{\Delta h_s}\frac{\varepsilon_{s,12}}{\varepsilon_{s,11}}k_{12}dx = \frac{1}{8}\left[\frac{3\varepsilon_{1,12}+\varepsilon_{2,12}}{\varepsilon_{1,11}}+\frac{\varepsilon_{1,12}+3\varepsilon_{2,12}}{\varepsilon_{2,11}}\right](\varepsilon_{1,12}-\varepsilon_{2,12}).
\end{equation}
Similarly, for other three terms in Eq. (\ref{18}),  we have
\begin{equation}\label{22}
\begin{aligned}
&\mathop{lim}\limits_{s\rightarrow 0}\int_{-\Delta h_s}^{\Delta h_s}\frac{\varepsilon_{s,1i}}{\varepsilon_{s,11}}k_{1j}dx = \frac{1}{8}\left[\frac{3\varepsilon_{1,1i}+\varepsilon_{2,1i}}{\varepsilon_{1,11}}+\frac{\varepsilon_{1,1i}+3\varepsilon_{2,1i}}{\varepsilon_{2,11}}\right](\varepsilon_{1,1j}-\varepsilon_{2,1j})\ (i,j=2,3).
\end{aligned}
\end{equation}
By substituting the four formulas above into $\int_{-\Delta h_s}^{\Delta h_s}Xdx$, the final components are calculated:
\begin{equation}\label{23}
  \begin{aligned}
  &\mathop{lim}\limits_{s\rightarrow 0}\int_{-\Delta h_s}^{\Delta h_s}X_{ij}dx = \Delta\left(\varepsilon_{ij}-\frac{\varepsilon_{1i}\varepsilon_{1j}}{\varepsilon_{11}}\right)\ (i,j=2,3).
  \end{aligned}
\end{equation}

To summarize, we transform Eq .\ref{14} to be
\begin{equation}\label{24}
  \mathop{lim}\limits_{s\rightarrow 0}\int_{-\Delta h_s}^{\Delta h_s}Xdx = \Delta \varepsilon_{11}^{-1}
  \begin{bmatrix}
    -1 & \varepsilon_{12} & \varepsilon_{13} \\
    \varepsilon_{12} &  \varepsilon_{11}\varepsilon_{22}-\varepsilon_{12}^2 &  \varepsilon_{11}\varepsilon_{23}-\varepsilon_{12}\varepsilon_{13} \\
    \varepsilon_{13} &  \varepsilon_{11}\varepsilon_{23}-\varepsilon_{12}\varepsilon_{13} &  \varepsilon_{11}\varepsilon_{33}-\varepsilon_{13}^2
  \end{bmatrix}
  = \xi
\end{equation}
Combining this equation with Eq. (\ref{11}), we derive a perturbation theory for calculating the eigenfrequency of anisotropic materials when shifts of boundaries occur.

\section{Simulation Verification}
In this section, we use the Finite Element Method (FEM) to analyze the accuracy of our formula, in comparison with the previous isotropic-material approximation scheme. 

\begin{figure}
\subfigure[]{
\begin{tikzpicture}[scale=1.35]
\coordinate (O) at (0,0,0);
\coordinate (A) at (0,2,0);
\coordinate (B) at (0,2,2);
\coordinate (C) at (0,0,2);
\coordinate (D) at (2,0,0);
\coordinate (E) at (2,2,0);
\coordinate (F) at (2,2,2);
\coordinate (G) at (2,0,2);

\draw[black, -stealth] (C) -- (3,0,2) node [right] {x};
\draw[black, -stealth] (C) -- (0,0,-1) node [right] {y};
\draw[black, -stealth] (C) -- (0,3,2) node [right] {z};
\draw[blue,fill=gray!20,opacity=0.4] (O) -- (C) -- (G) -- (D) -- cycle;
\draw[blue,fill=gray!20,opacity=0.4] (O) -- (A) -- (E) -- (D) -- cycle;
\draw[blue,fill=gray!20,opacity=0.4] (O) -- (A) -- (B) -- (C) -- cycle;
\draw[blue,fill=blue!40,opacity=0.4] (D) -- (E) -- (F) -- (G) -- cycle;
\draw[blue,fill=gray!20,opacity=0.4] (C) -- (B) -- (F) -- (G) -- cycle;
\draw[blue,fill=gray!20,opacity=0.4] (A) -- (B) -- (F) -- (E) -- cycle;
\draw[|<->|] (2.1,0,2) -- node [right] {100 nm} (2.1,0,0);
\draw[|<->|] (2.1,0,0) -- node [right] {100 nm} (2.1,2,0);
\draw[|<->|] (2,2.1,0) -- node [above] {100 nm} (0,2.1,0);
\draw[fill] (C) circle (1pt);
\node[below] at (0,0,2) {O};
\draw[->,very thick,red] (2,1,1) -- (2.7,1,1);
\end{tikzpicture}
\label{fig.41}
}
\quad
\subfigure[]{
\includegraphics[width=5cm, height=5cm]{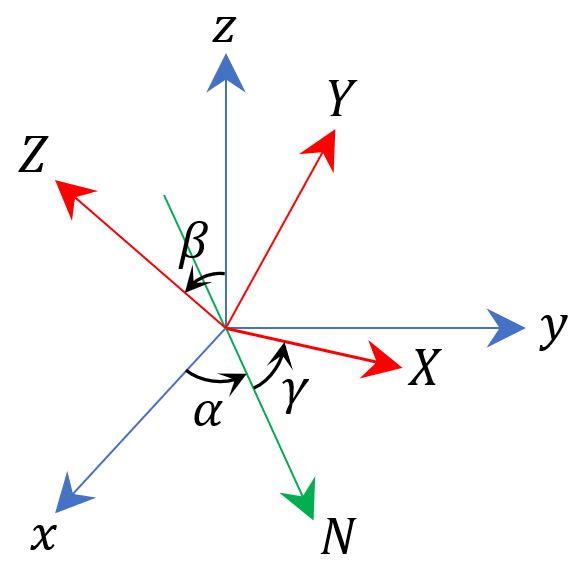}
\label{fig.42}
}
\caption{(a) The cube resonant cavity with the length of each edge L=100 nm. The shifts of boundaries are imposed on the blue color side. The red arrow indicates the orientation of the shifts. Definition of Euler angles $\alpha,\ \beta,$ and $\gamma$ is shown in (b), where the axes $X$, $Y$, and $Z$ indicate three principal axes of the index ellipsoid of the cavity material.}
\label{fig.4}
\end{figure}
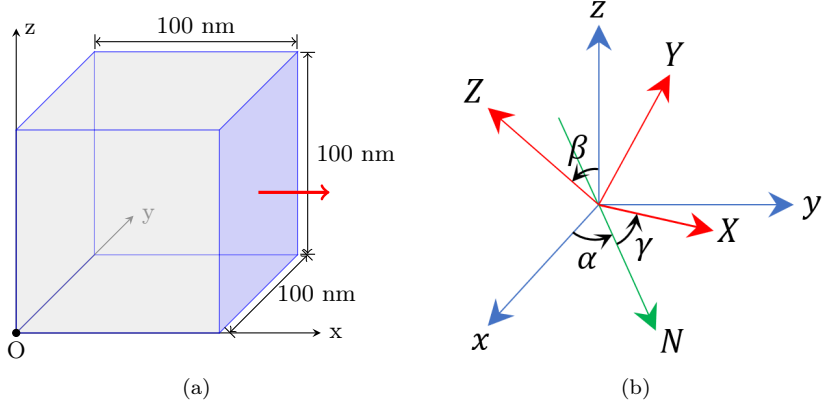

We consider a resonant cavity and calculate its eigenfrequency derivative as a function of deformation through both our and previous methods. Here we choose a cubic cavity with the length of each edge $L=100\ nm$ surrounded by air, as depicted in Fig.\ref{fig.41}. The computational domain is a cube with the length of each edge to be $5L$, and all sides of the computational domain cube are parallel to the corresponding sides of the resonant cavity cube, while the cavity locates in the center of the cube. We set the first-order scattering  boundary condition on all interfaces of the computational domain. Dielectric loss is ignored and the refractive indexes of the cavity material are set to be $n_o=10,\ n_e=9$. With such high refractive indexes, we can tighten the distribution of electromagnetic field so as to reduce the amount of calculation. As a comparison, we set the refractive index of the cavity to be $n=9.5$ for the isotropic approximation situation, following a recent work on Lithium Niobate \cite{1903}. The Euler angles $\alpha$, $\beta$, $\gamma$ corresponding to the rotation angles of three spindles $X$, $Y$, and $Z$ of the index ellipsoid are defined with basis vectors in the space coordinate system (see Fig.\ref{fig.42}). The displacement imposed on the blue side $\Delta x_q(y,z)=\frac{q}{2\pi{\sigma}^2}exp\left(-\frac{y'^2+z'^2}{2\sigma^2}\right)$ is perpendicular to that side of the cavity, where $\sigma=20,\ y'=y-50,\ z'=z-50$, and the orientation of the displacement is indicated by the red arrow in Fig.\ref{fig.41}. (Here we omit the common unit of 'nm'.) To simplify, here we investigate the case where $\beta=\gamma=0$, $\alpha\in[0,\frac{\pi}{2}]$ and the frequency of the investigated mode is about 360 THz. The parameter of the perturbation is set as $q=1000$, when the maximum shift of 0.4 nm occurs at the center of the blue side.

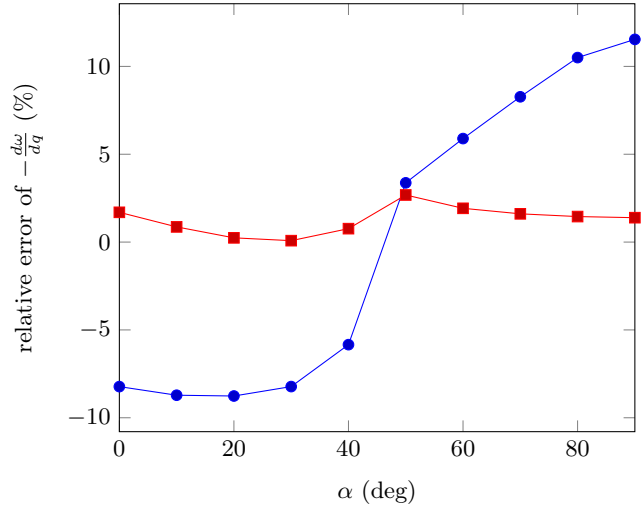
\begin{figure}
\begin{tikzpicture}
\begin{axis}[
    xlabel = {$\alpha\ $(deg)},
    ylabel = {relative error of $-\frac{d\omega}{dq}$ (\%)},
    xmin=0, xmax=90,
    legend pos=south east,
    legend style ={ at={(0.2,0.02)},
    anchor=south west,
    draw=black,
    fill=white,
    align=left,
    },
]

\addplot coordinates {
    (0,-0.0822675*100)(10,-0.087166208*100)(20,-0.087616874*100)(30,-0.082247957*100)(40,-0.058388252*100)(50,0.033726707*100)(60,0.058916557*100)(70,0.082697545*100)(80,0.105016666*100)(90,0.115378865*100)
    };

\addplot coordinates {
    (0,0.016946413*100)(10,0.008643751*100)(20,0.002415422*100)(30,0.000807702*100)(40,0.007643252*100)(50,0.026808448*100)(60,0.019223523*100)(70,0.016084863*100)(80,0.014525587*100)(90,0.013884096*100)
    };

\end{axis}
\end{tikzpicture}
\caption{Estimated relative errors for $\frac{d\omega}{dq}$, as a function of $\alpha$. Here $\beta=\gamma=0$. The results of our anisotropic perturbation theory and previous isotropic method are in red and blue, respectively.}
\label{fig.5}s
\end{figure}

In order to compare the accuracy between our method and previous isotropic-approximation scheme, we first calculate the eigenfrequency derivative as a function of deformation $\frac{d\omega}{dq}$ by numerical method, which will be considered as the ‘true’ value. Then we calculate $\frac{d\omega}{dq}$ by our 'anisotropic' method and previous isotropic-approximation scheme, respectively. Compared with the 'true' value, the estimated relative errors are shown in Fig.\ref{fig.5}, as a function of $\alpha$. It is shown that our method yields much smaller errors in most cases. 

\section{Discussion and conclusion}
In Fig.\ref{fig.3}, we have shown three smoothing functions, and $g_s(x)$ and $f_{s0}(x)$ are used for smoothing different components. Why not directly choose a rectangular smoothing function $f_{s0}(x)$ to simplify the derivation? Actually, by using a rectangular smoothing function, one can derive the same result. However, it seems that there is an assumption that all physical smoothing methods will lead to the same result. Conversely, our derivation does not rely on such an assumption. 

Here, 'physical' means that the smoothing should have physical meanings. For example, the imposed restrictions on the smoothing that the smoothed permittivity tensor field should be symmetric positive-defined means that the smoothed permittivity field may correspond to some real matter. 

A few factors that would cause frequency shifts are not taken into account, such as contribution from the edges. Johnson et al. \cite{Johnson2002} have demonstrated that the influence from factors except the shifts of sides are negligible for 1-order approximation, which guarantees the accuracy of our results.

Our perturbation theory incorporates Johnson's scheme. In the case of isotropic permittivity, if we denote $\varepsilon_1 - \varepsilon_2$ by $\Delta \varepsilon$, then it will yield $\xi = \Delta Diag\{-\varepsilon^{-1}, \varepsilon, \varepsilon\}$. In this way our formula is the same as Johnson's result\cite{Johnson2002} under isotropic conditions. 

Our anisotropic perturbation theory can help in many situations. For example, Lithium Niobate optomechanical crystals have several promising applications like ultra-low-power modulators and quantum information processing, and a key step of designing these crystals is to calculate the single-phonon optomechanical coupling rate in anisotropic materials\cite{1903}. The optomechanical coupling rate includes two terms, one term comes from the change of the crystal permittivity by pressure and the other term originates from the shifts of crystal interfaces. The connection is, the calculation of the second term involves the calculation of eigenfrequency's derivative with respect to deformation, which needs Johnson's or our perturbation method. While previous isotropic-approximation method causes errors in the second term for anisotropic situations, our method can almost eliminate the errors. Using our scheme to calculate the second term, the coupling rate is
\begin{equation}\label{25}
g_{0,MB} = -\frac{\omega^{(0)}}{2}\frac{\int\limits_{S}dA [D_x,E_y,E_z]^* \xi [D_x, E_y, E_z]' h}{\int E\cdot D^* dV},
\end{equation}
where $h$ is the displacement of the interface in $x$-axis orientation for mechanical vibration. The coordinate system is defined in Fig.\ref{fig.1}, and $\xi$ is defined in Eq. (\ref{24}). The modification of eigenfrequency derivative based on our anisotropic theory is typically $\Delta g_{0,MB} = 10$ kHz, while the total coupling rate is several hundred kHz.

Nevertheless, the calculation of perturbed field modes is still elusive as the expansion of $E^{(1)}$ by $E^{(0)}$ (or similar expansion of magnetic field) may fail as the shifts of boundaries\cite{Johnson2002}. This is, in fact, a fundamental theoretical problem of perturbation theory. Another problem is how to develop the perturbation theory of Maxwell's equations dielectric loss is taken into account. A possible method is to transform the form of Maxwell's equations to construct a real and symmetric coefficient tensor. More theoretical work is still needed in this field.

In conclusion, we report a perturbation theory of Maxwell's equations for anisotropic dielectric interfaces. This theory shows a better accuracy compared with previous isotropic-approximation method. Our method could play an important role in the future applications of anisotropic materials. 

\begin{acknowledgments}
This work is partially supported by National Key Research and Development Program of China (2018YFA0306102, 2018YFA0307400, 2017YFA0304000); National Natural Science Foundation of China (91836102, 61704164, 12074058, 61775025, 61705033, 61405030, 61308041).
\end{acknowledgments}

\end{document}